# FeTe$_{1-x}$Se$_x$ monolayer films: towards the realization of high-temperature connate topological superconductivity

X. Shi[1†], Z.-Q. Han[1,2†], P. Richard[1,3,4], X.-X. Wu[1], X.-L. Peng[1], T. Qian[1,3], S.C. Wang[2], J.P. Hu[1,3,4], Y.-J. Sun[1*] & H. Ding[1,3,4*]

[1]*Beijing National Laboratory for Condensed Matter Physics, and Institute of Physics, Chinese Academy of Sciences, Beijing 100190, China*

[2]*Department of Physics, Renmin University, Beijing, 100872, China*

[3]*Collaborative Innovation Center of Quantum Matter, Beijing 100190, China*

[4]*School of Physical Sciences, University of Chinese Academy of Sciences, Beijing 100190, China*

[†]*These authors contributed equally to this work*

**[Abstract] We performed angle-resolved photoemission spectroscopy studies on a series of FeTe$_{1-x}$Se$_x$ monolayer films grown on SrTiO$_3$. The superconductivity of the films is robust and rather insensitive to the variations of the band position and effective mass caused by the substitution of Se by Te. However, the band gap between the electron- and hole-like bands at the Brillouin zone center decreases towards band inversion and parity exchange, which drive the system to a nontrivial topological state predicted by theoretical calculations. Our results provide a clear experimental indication that the FeTe$_{1-x}$Se$_x$ monolayer materials are high-temperature connate topological superconductors in which band topology and superconductivity are integrated intrinsically.**

High-temperature superconductors, hosts to a non-conventional electron pairing mechanism, and topological electronic materials, for which exotic metallic surface states are protected by symmetries, are two intensively studied areas. Plenty of efforts have been made to search for topological superconductors [1, 2]. The simplest model deals with an unconventional chiral superconductor (Fig. 4(c) upper panel), but there is still no convincing experimental evidence [3, 4]. The second type of popular proposals is based on the proximity effect in heterostructures made of topological insulators and *s*-wave superconductors (Fig. 4(c) middle panel). However, this needs relatively large coherence length and the superconducting transition temperature ($T_c$) is usually low [5, 6].

Interestingly, there is a third type of proposals for realizing topological superconductors in superconductors with topologically nontrivial bands. Here we refer to this type of realization as connate topological superconductivity due to the intrinsic integration of band topological properties and superconductivity. Some iron-based superconductors have been proposed to fulfill this type of realization. For example, the monolayers of FeSe grown on $SrTiO_3$ (STO) substrates, which have a maximum $T_c$ higher than 50 K [7, 8, 9, 10, 11, 12, 13], may have a nontrivial topology due to band inversion at the Brillouin zone boundary (M point) [14, 15]. However, this band inversion is always far below the Fermi surface as these systems are electron doped. Another candidate is the $FeTe_{1-x}Se_x$ system, in which a topologically nontrivial band inversion takes place near Fermi surfaces around the Z point [16]. The shortcoming of these systems is that the maximum $T_c$ is only about 14 K [17, 18].

In this paper, we report that $FeTe_{1-x}Se_x$/STO(001) monolayers have a high $T_c$ similar to that of FeSe/STO monolayers and simultaneously can carry topologically nontrivial band inversion as the bulk $FeTe_{1-x}Se_x$. Thus, this system is a possible candidate to realize high-temperature connate topological superconductivity. We substitute Se by Te in FeSe/STO monolayer films and carry out a systematic angle-resolved photoemission spectroscopy (ARPES) investigation on a series of films with different Se concentrations. The Fermi surface (FS) results suggest comparable high charge transfer in all samples. By tracing the temperature dependence of the superconducting gap, we show that superconductivity persists upon Te substitution of Se, in spite of the dramatic variations of the electronic structure. The electron- and hole-like bands at the Brillouin zone (BZ) center Γ shift towards each other and give rise to a rapid decrease of the band gap, resulting in an evolution towards a band inversion and nontrivial $Z_2$ topological invariant. Our results suggest a highly possible topological phase transition occurring at a low Se concentration in FeTe$_{1-x}$Se$_x$ monolayer films. We propose several ways to search for topological superconductivity in this unique system by taking advantage of its high flexibility and fine tunability.

Monolayer films of FeTe$_{1-x}$Se$_x$ were epitaxially grown on 0.05wt% Nb-doped STO(001) substrates by the same process as FeSe/STO(001) described in our previous work [12]. We have grown samples with different Se concentrations by tuning the flux ratio between Se (99.999%) and Te (99.99%). The Te/Se ratios of the films were estimated based on the Vegard's law and the linear evolution of the band positions at Γ, which will be discussed below, to give the nominal Se concentrations. This can be deviated from true value but will not affect our conclusion. Then the films were transferred *in situ* into the ARPES chamber after annealing. ARPES measurements were carried out using a R4000 analyzer and a helium discharge lamp under ultrahigh vacuum better than $3 \times 10^{-11}$ torr. The data were recorded with He Iα photons ($h\nu$ = 21.218 eV). The energy resolution was set to ~ 5 meV for gap measurements and ~ 10 meV for the band structure and FS mapping. The angular resolution was set to 0.2°.

The FeSe/STO monolayer films possess a simple FS topology characterized by large electron-like pockets at the M point and the absence of hole-like pockets at Γ [8, 9, 12], which differs from that of FeSe bulk materials and most of the iron-pnictide superconductors [19]. We show in Fig. 1(a) the evolution of the FS of FeTe$_{1-x}$Se$_x$/STO monolayer as a function of the Se content. The electron pockets at M can be observed in all samples. The size of the pockets is rather independent of the substitution, and leads to a carrier concentration of about 0.16 electrons per unit cell according to the Luttinger theorem. This indicates that the charge transfer in this heterostructure is robust, regardless of the Se concentration. We note that the coherence of the spectrum degrades upon Te substitution, which may be due to enhanced electron correlations

[20, 21] or antiferromagnetic fluctuations [22]. Strong intensity appears at the BZ center as the Se decreases. We will explain below that this originates from an electron band in the unoccupied side, similar to the case of K deposited FeSe/STO monolayer [12] and K deposited bulk FeTe$_{1-x}$Se$_x$ [23].

As the FeTe$_{1-x}$Se$_x$/STO monolayer films share similar FS topology, which may be a key ingredient for the high superconducting transition temperature in FeSe monolayers [12], we check the evolution of superconductivity in this system. Figure 1(b) displays the temperature dependence of the symmetrized energy distribution curves (EDCs) at the Fermi momentum ($k_F$) point of the electron FS around M. Although the introduction of Te weakens the superconducting coherence peak, a gap feature can be clearly observed at the lowest temperature for each sample. Both the gap sizes and the closing temperatures have similar magnitude. A closer inspection shows that the gap is a little larger for the samples in which additional intensity of FS appears at Γ. These results are consistent with the enhancement of superconductivity accompanying a Lifshitz transition in electron-doped FeSe monolayer [12].

We then address the electronic band structure in detail. The electron-like bands observed around the M point are displayed in Fig. 2(a). The spectra are divided by the Fermi-Dirac (FD) function convoluted by a Gaussian resolution function. We fit the band dispersions using parabolic curves and investigate the evolution in FeTe$_{1-x}$Se$_x$/STO. The band bottom (Fig. 2(b)) moves towards the Fermi level ($E_F$) upon decreasing the Se concentration, while the effective mass (Fig. 2(c)) increases from FeSe/STO to FeTe/STO, which is easy to understand considering the strong electronic correlations in the FeTe bulk material [24]. As a result, the $k_F$ value or FS size does not change much, as shown in Fig. 2(d).

Figure 3 displays the band evolution around Γ of FeTe$_{1-x}$Se$_x$/STO. We observe dramatic variations compared to that at the M point. The curvature plots in Fig. 3(b), which can highlight the band dispersions, show two hole-like bands carrying a $d_{xz}/d_{yz}$ orbital character. We now examine the inner one, which is also closer to $E_F$. The band top moves up from FeSe to FeTe, similarly to the electron band bottom at M. In contrast, the effective mass evolves in the opposite way (Fig. 3(f)). We show in Fig. 3(a) the spectra after division by the FD function in order to check the unoccupied band. A down-shifting electron-like band is clearly observed when the Se content is smaller than 60%, similar to the case of FeTe$_{1-x}$Se$_x$ single crystals [23]. This electron-like band, which has a $p_z$ orbital character hybridized with $d_{xy}$ in the monolayer FeSe/STO, according to LDA calculations [25, 26]. The electron- and hole-like bands move towards each other and the band gap between them decreases rapidly. Eventually the bands touch each other at a Se concentration of approximately 33%, which is further revealed in the plots of the constant energy contours and momentum distribution curves (MDCs), as shown in Fig. 3(c)-(d). Since the bands deviate from the parabolic dispersion towards the FeTe side, we fit the bands to the form $E = C_0 + C_1|k| + C_2 k^2$ and compare the results in Fig. 3(e). The more linear-like trend of the band dispersion is a consistent feature of the bands towards band touching.

Our main results on FeTe$_{1-x}$Se$_x$/STO monolayers are summarized in Fig. 4. Superconductivity persists in this system (Fig. 4(a)), which is consistent with scanning tunneling microscopy (STM) results [27], but the electronic structure undergoes significant evolution. Firstly, all the samples share a similar FS topology, which is mainly characterized by the large electron pockets at the M point. This FS with high electron concentration is believed to favor superconductivity at high temperature [12]. Secondly, the increasing effective mass of the band at M could be attributed to the increasing correlation from FeSe to FeTe, while the opposite evolution of that at Γ is possibly due to the coupling with the $p$ orbitals of Se/Te, which is a consequence of the band shift towards a band touching. These effects seem to have no link to the $T_c$. Thirdly, the band shifting at the large energy scale does not affect superconductivity either, except for a $T_c$ enhancement accompanying the emergence of the $d_{xy}/p_z$ band near $E_F$, which has been discussed elsewhere [12].

The most significant change upon Te substitution is that the decrease of the band gap at Γ may trigger a topological phase transition. The hole- and electron-like bands touch each other at a Se concentration of about 33% (Fig. 4(b)). Theoretical calculations on single layer FeTe$_{1-x}$Se$_x$ [25] give similar evolution of the band structure, and it was demonstrated that the band inversion and parity exchange at Γ, as illustrated in Fig. 4(b), will lead to a nontrivial $Z_2$ topological invariant at the FeTe side. In fact, a topologically nontrivial phase in FeSe$_{0.5}$Te$_{0.5}$ bulk crystal originating from the band inversion along the Γ-Z direction was also found recently by density functional theory and ARPES studies [16]. A following theoretical analysis predicted that a topological superconducting phase can be realized on the (001) surface of this material [28].

The monolayer films that we studied have distinct advantages compared to the bulk materials, such as a higher $T_c$ and an easy-tunable electronic structure. The latter is especially essential for driving the topological phase transition and the tuning can be achieved by plenty of ways of interfacial engineering, including the choice of substrate and orientation [13], element substitution like Se by Te in our case, and surface perturbation like K deposition [12]. All these methods have been proved to be highly effective. Therefore, the FeTe$_{1-x}$Se$_x$ monolayer system becomes an ideal platform for combining high-temperature superconductivity and nontrivial topological properties. We anticipate more direct evidence in future studies, such as topological edge states, which could be detected by scanning tunneling microscopy, or by ARPES on a sample grown on a substrate with a high density of steps.

Our studies provide initial experimental evidence for realizing topological superconductivity with relatively high transition temperature in the monolayer FeTe$_{1-x}$Se$_x$ system. As sketched in Fig. 4(c), this provides a new path to the condensed-matter realization of Majorana zero modes other than previous proposals such as searching in *p*-wave superconductors and the vortices in the Fu-Kane model (proximity effect between a topological insulator and a s-wave superconductor) [1, 2, 3, 4, 5], which are difficult to achieve with a high transition temperature due to either intrinsic low-$T_c$ or short coherent length of high-$T_c$ superconductors. In contrast, this new system (connate topological superconductor) divides the "working responsibilities" of superconductivity and topological band inversion between different bands sharing the superconducting quantum coherence among them. As a consequence, it could achieve high-temperature topological superconductivity in a single phase, which is highly desirable for application purposes.

**Acknowledgements**

We thank Zhenyu Zhang for useful discussions. We also thank Yunlong Guo for assistance in ARPES measurements. This work is supported by grants from the Ministry of Science and Technology of China (2015CB921000, 2016YFA0401000, 2015CB921301) and the National Natural Science Foundation of China (11574371, 11274362, 1190020, 11334012, 11274381).

**Additional information**

The authors declare no competing financial interests. Correspondence and requests for materials should be addressed to Y.-J.S. (yjsun@iphy.ac.cn) and H.D. (dingh@iphy.ac.cn).

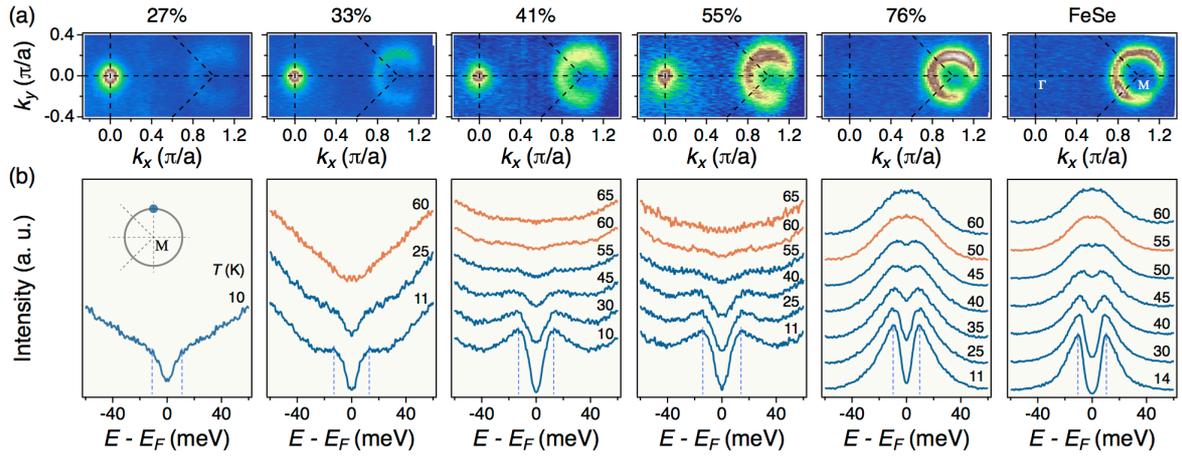

Figure 1 (a) ARPES FS maps of a series of FeTe$_{1-x}$Se$_x$ monolayers grown on STO, with the nominal concentration of Se indicated above each panel. (b) Temperature evolution of the symmetrized EDCs at the $k_F$ point of the electron FS around M for each sample. The orange curves correspond nearly to the gap closing temperature.

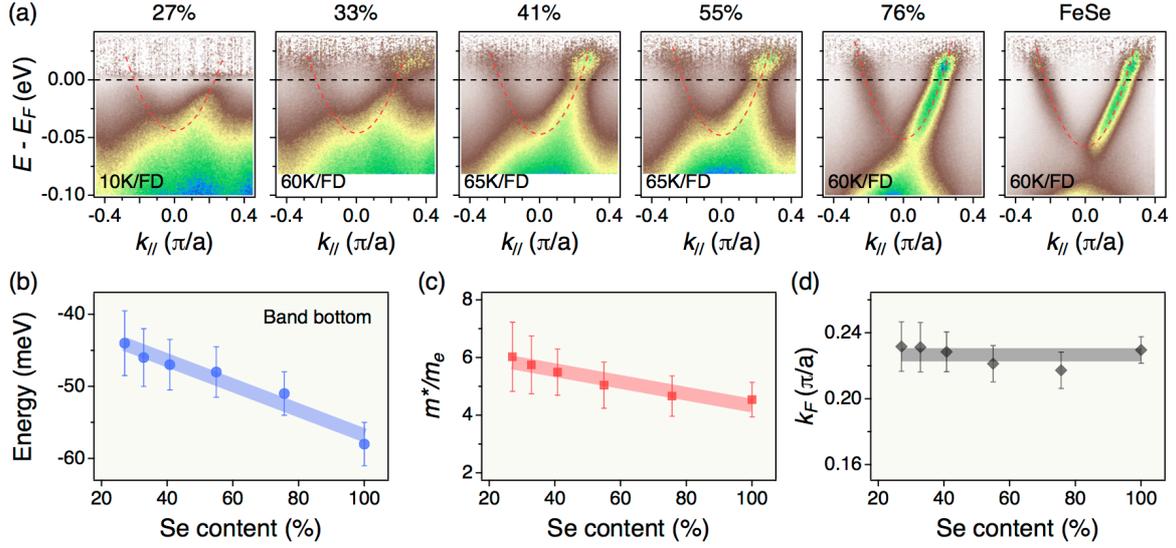

Figure 2 (a) Evolution of the ARPES intensity plot near M along the Γ-M direction for 1UC FeTe$_{1-x}$Se$_x$/STO. The plots are divided by the Fermi-Dirac distribution function to visualize the states above $E_F$. The red curves are parabolic fits to the band dispersions. (b)-(d) Se concentration dependence of the band bottom, effective mass and Fermi momentum of the band shown in (a), respectively.

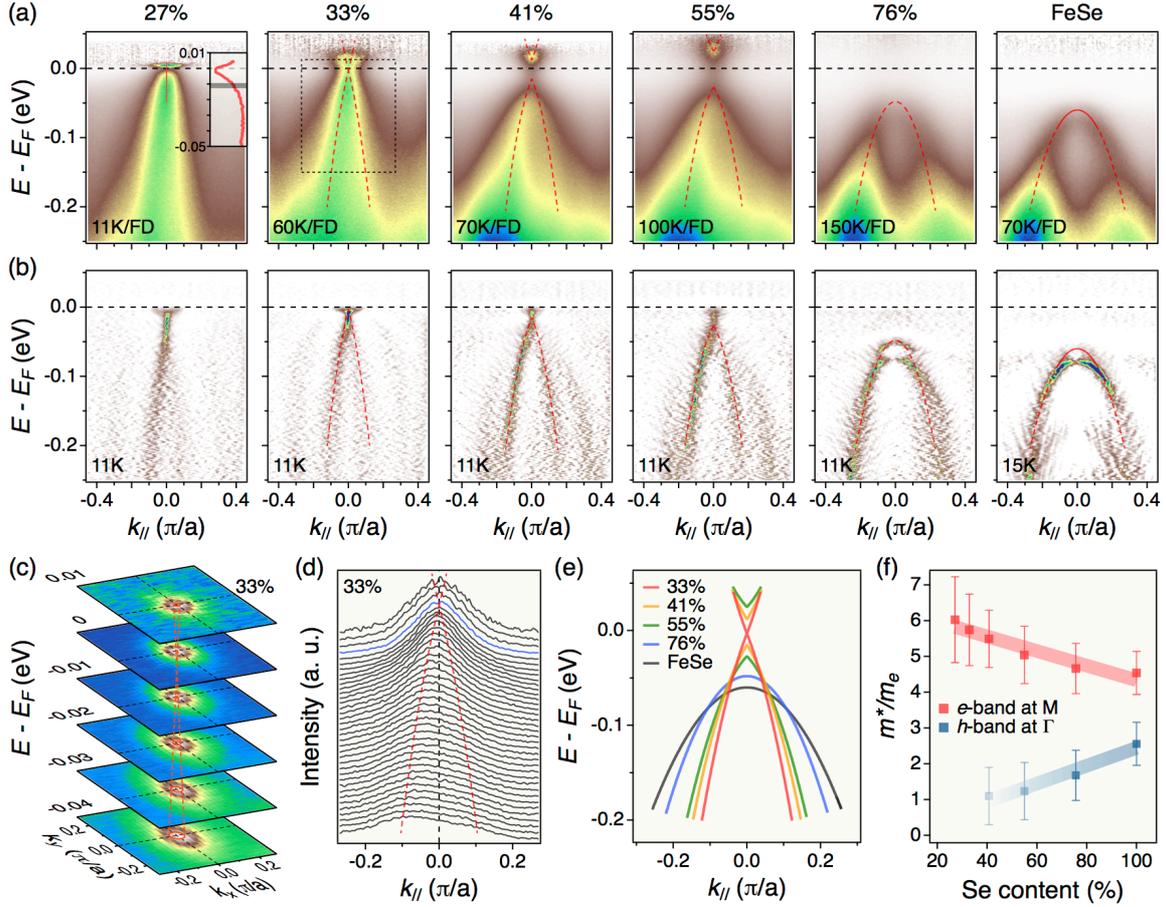

Figure 3 (a) Evolution of the intensity plot divided by the Fermi-Dirac distribution function near Γ along the Γ-M direction for 1UC FeTe$_{1-x}$Se$_x$/STO. The inset of the Se-27% panel displays the EDC at Γ and shows a gap-like feature. (b) Curvature intensity plots along the same cut as in (a). The data were recorded at the temperature indicated in each panel. The red curves fit the band dispersion in the form of $E = C_0 + C_1|k| + C_2 k^2$. (c) Intensity plot of the constant energy contours at different binding energies for the Se33% sample. (d) MDC plot corresponding to the spectrum in the black square in (a). (e) Comparison of the band dispersions at Γ for each sample. (f) Se concentration dependence of the effective mass. The results of the electron band at M shown in Figure 2(c) are included for a comparison.

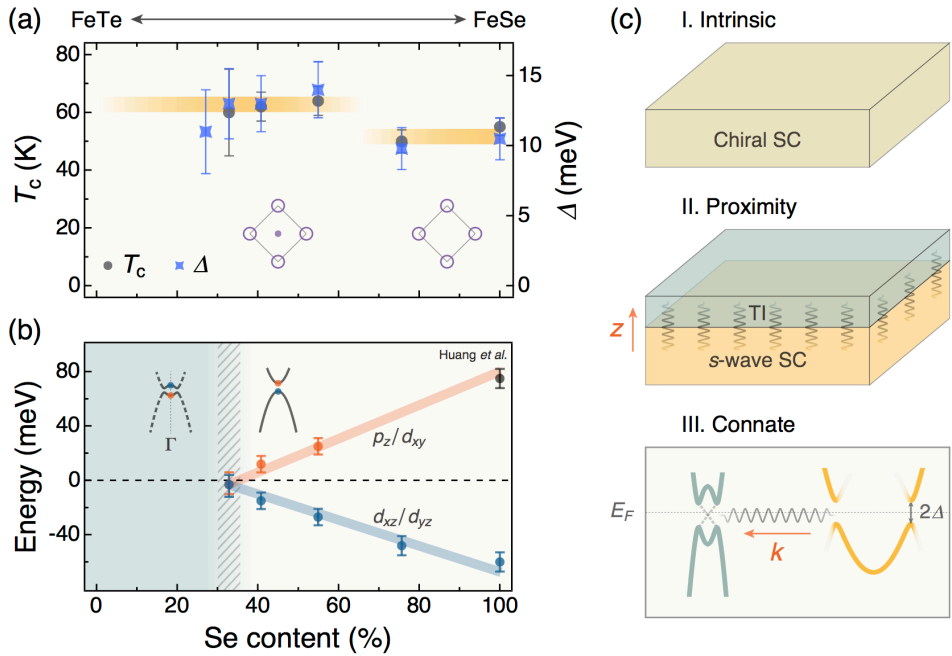

Figure 4 (a) Phase diagram illustrating the evolution of the superconducting gap and its closing temperature in 1UC FeTe$_{1-x}$Se$_x$/STO. Two typical FS topologies are also plotted in the diagram. (b) Evolution of the band positions at Γ, showing a trend towards band inversion upon decreasing the Se concentration. The electron-like band position of FeSe/STO is adapted from Ref. [26]. (c) Schematic plots of different types of method to realize topological superconductor.